\documentclass{article}  
\usepackage{jamaica04}
\usepackage[dvips]{graphicx}
\frompage{000} \topage{000}                                              

\def\rightmark{Hyperon production at CERN SPS energies}
\def\leftmark{Michael Mitrovski}
 
\title{Hyperon production at CERN SPS energies} 
\authors{{Michael Mitrovski$^{1}$ for the NA49 Collaboration$^{a}$
}\\[2.812mm]
{\normalsize
\hspace*{-8pt}$^1$ Insitut f\"ur Kernphysik, August-Euler-Strasse 6\\ 
60486 Frankfurt, Germany\\ \\
{\bf E-mail: Michael.Mitrovski@cern.ch} 
}}

\abstract{New experimental results of NA49 on hyperon production in central Pb+Pb collisions at the SPS energies are presented. In particular, measurements of $\Lambda$ production at 30 A$\cdot$GeV and $\Xi$ and $\Omega$ production at 40 A$\cdot$GeV are shown. Transverse mass spectra and rapidity distributions of hyperons at different energies are compared. The energy dependence of the particle yields and ratios is discussed.}

\PACS{specifications see, e.g.\ {\tt http://www.aip.org/pacs/}}
 
\begin{document}
 
\maketitle
\setcounter{page}{1}

\section{Introduction}
\label{Introduction}

A non-monotonic energy dependence of the $K^{+}$/$\pi^{+}$ ratio with a sharp maximum close to 30 A$\cdot$GeV is observed in central Pb+Pb collisions  \cite{A.1}. Within the statistical model of the early stage \cite{A.2}, this is interpreted as a signal of the phase transition to a QGP, which causes a sharp change in the energy dependence of the strangeness to entropy ratio. This observation naturally motivates further study of the production of hyperons as a function of the beam energy. \\
Furthermore it was suggested that the kinematic freeze-out of $\Omega$ takes place directly at QGP hadronization. If this is indeed the case, the transverse momentum spectra of the $\Omega$ should directly reflect the transverse expansion velocity of a hadronizing QGP \cite{A.3,A.4}. \\   
In this report we show the new NA49 results on hyperon production at central Pb+Pb collisions at the low SPS energies and compare them to previously shown data at other energies.

\newpage 

\section{The NA49 Experiment}\label{NA49_Experiment}  

\begin{figure}[h]
\begin{center}
\includegraphics[scale=0.38]{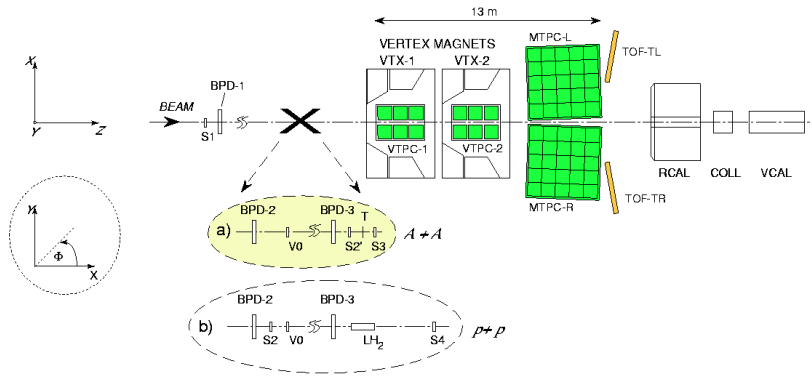}   
\caption{The NA49 experimental setup.}  
\label{fig:Experiment}
\end{center}
\end{figure}

The NA49 detector \cite{A.5}, shown in Fig.~\ref{fig:Experiment}, is a large acceptance hadron spectrometer at the CERN SPS, consisting of four large volume TPCs. Two of them, the Vertex TPCs (VTPC), are inside a magnetic field for the determination of particle momenta and charge. The ionisation energy loss (dE/dx) measurements in the two Main TPCs (MTPC), which are outside the magnetic field, are used for particle identification.
 \vspace{-0.5cm}
Central collisions were selected by triggering on information from a downstream calorimeter (VCAL),  which measures the energy of the projectile spectator nucleons. 

\section{Inverse slope parameters and transverse mass spectra}

\begin{figure}[hbt!]
\begin{center}
\includegraphics[scale=0.45]{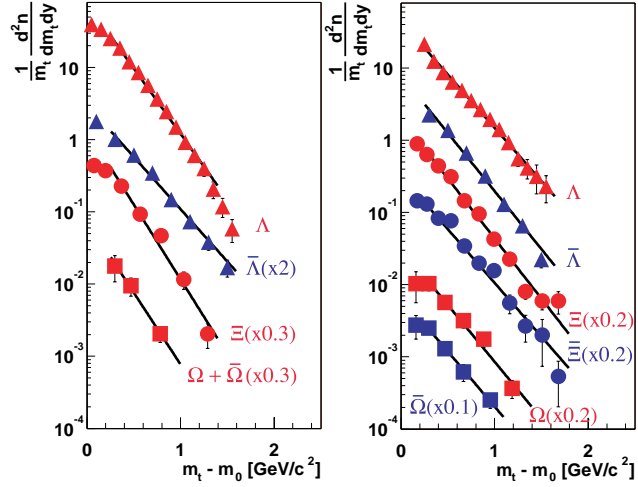} 
\caption{Transverse mass spectra of $\Lambda$ (triangles), $\Xi$ (circles) and $\Omega$ (squares) and their antiparticles at midrapidity in central Pb+Pb collisions at 40 (left) and 158 A$\cdot$GeV (right). The spectra of the $\Lambda$ hyperons are not corrected for feeddown from $\Xi$ and $\Omega$ decays.}  
\label{fig:mt_spectra} 
\end{center}
\end{figure}

Fig.~\ref{fig:mt_spectra} shows the transverse mass ($m_{t}$ $=$ $\sqrt{p_{t}^{2} + m^{2}}$) spectra of the hyperons and antihyperons at 40 and 158 A$\cdot$GeV at midrapidity in central Pb+Pb collisions. At 40 A$\cdot$GeV the signal to background ratio is better than at 158 A$\cdot$GeV. Therefore it was possible to make measurements also in the low $m_{t}$-region. The distributions are fitted by the exponential function :
\begin{equation}
\frac{1}{m_{t}} \frac{d^{2}N}{dm_{t}dy} = C \cdot e^{- m_{t}/T} , 
\end{equation}
where the fit parameters are the normalization factor $C$ and the inverse slope parameter $T$. The fitted range is $m_{t}$ - $m_{0}$ $>$ 0.25 GeV/$c^{2}$ for all particles. The spectra at 158 A$\cdot$GeV are well described by the exponential function. At 40 A$\cdot$GeV the exponential fit does not describe the data in the low $m_{t}$-range. The resulting inverse slope parameters are summarized in Table~\ref{tab:Lambda_Table} - \ref{tab:Omega_Table}. \\
\begin{figure}[hbt!]
\begin{center}
\includegraphics[scale=0.37]{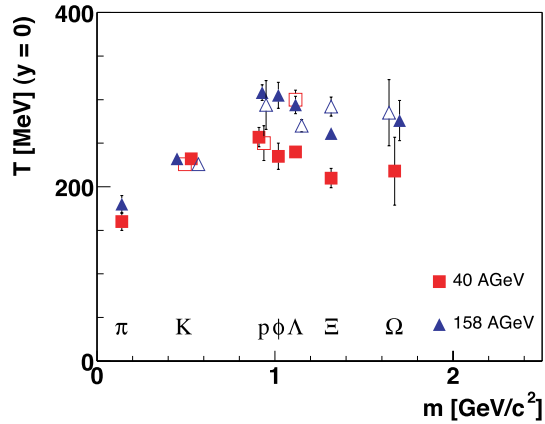}  
\caption{The inverse slope parameter as a function of the particle mass in central Pb+Pb collisions at 40 (squares) and 158 (triangle) A$\cdot$GeV. The results for antiparticles are indicated by the open symbols.}  
\label{fig:Nu_Xu_Plot}
\end{center}
\end{figure} \\
The results for $\Lambda$ at both energies~\cite{A.6} and for the $\Xi$ at 158 A$\cdot$GeV were published~\cite{A.7} already. All other results are still preliminary. A compilation of all NA49 data on the inverse slope parameter as a function of the particle mass in central Pb+Pb collisions at 40 and 158 A$\cdot$GeV is shown in Fig.~\ref{fig:Nu_Xu_Plot}. The inverse slope parameter increases with particle mass up to m $\approx$ 1 GeV/$c^{2}$. For heavier particles the $T$ parameter is approximately independent on the mass. This can be interpreted as a sign of early freeze-out of multistrange hyperons \cite{A.8}. At these high masses the $T$ parameter is by about 50 MeV lower at 40 A$\cdot$GeV than at 158  A$\cdot$GeV. 

\section{Rapidity spectra and total yields}

\begin{figure}[hbt!]
\begin{center}
\includegraphics[scale=0.6]{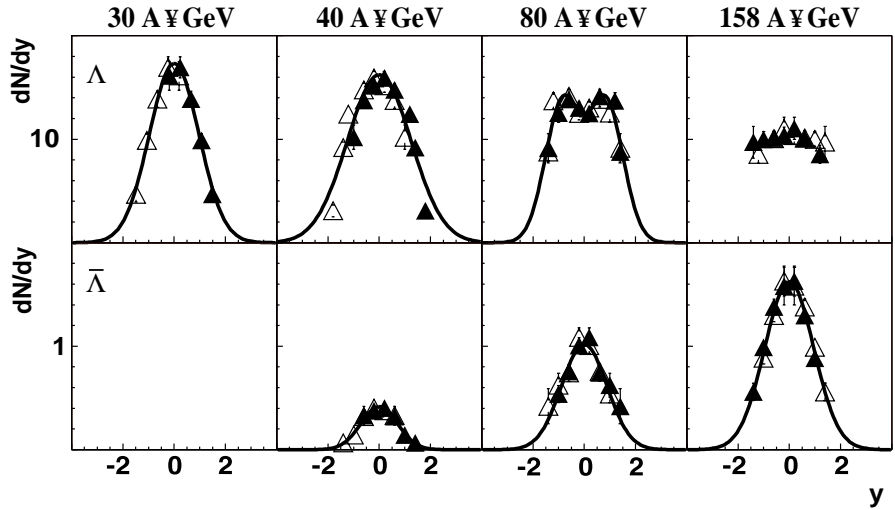}  
\caption{The rapidity spectra of $\Lambda$ (top) and $\bar{\Lambda}$ (bottom) in central Pb+Pb collisions at 30 to 158 A$\cdot$GeV. The full symbols are the measured points and the open points show their reflection with respect to midrapidity.}  
\label{fig:Rapidity_Spectra_Lambda}
\end{center}
\end{figure}

\begin{figure}[hbt!]
\begin{center}
\includegraphics[scale=0.55]{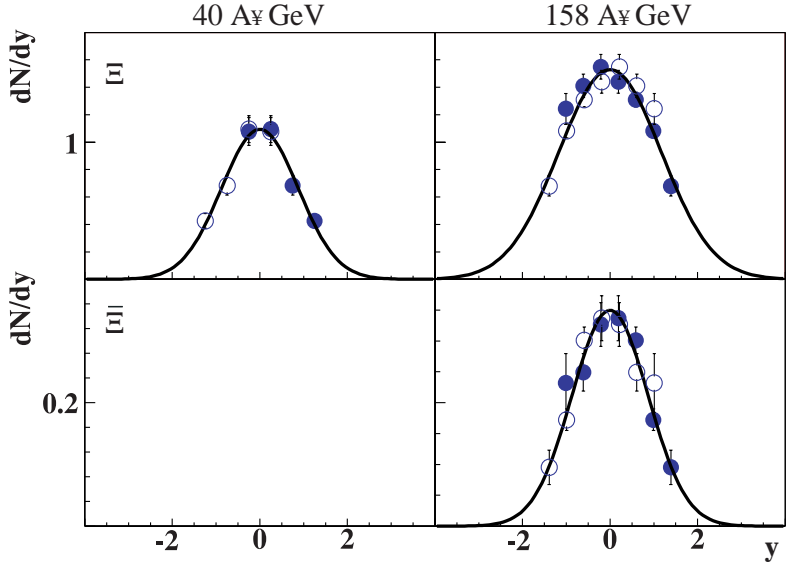} 
\caption{The rapidity spectra of $\Xi$ (top) and $\bar{\Xi}$ (bottom) in central Pb+Pb collisions at 40 and 158 A$\cdot$GeV. The full symbols are the measured points and the open points show their reflection with respect to midrapidity.} 
\label{fig:Rapidity_Spectra_Xi}
\end{center}
\end{figure}

\begin{figure}[hbt!]
\begin{center}
\includegraphics[scale=0.48]{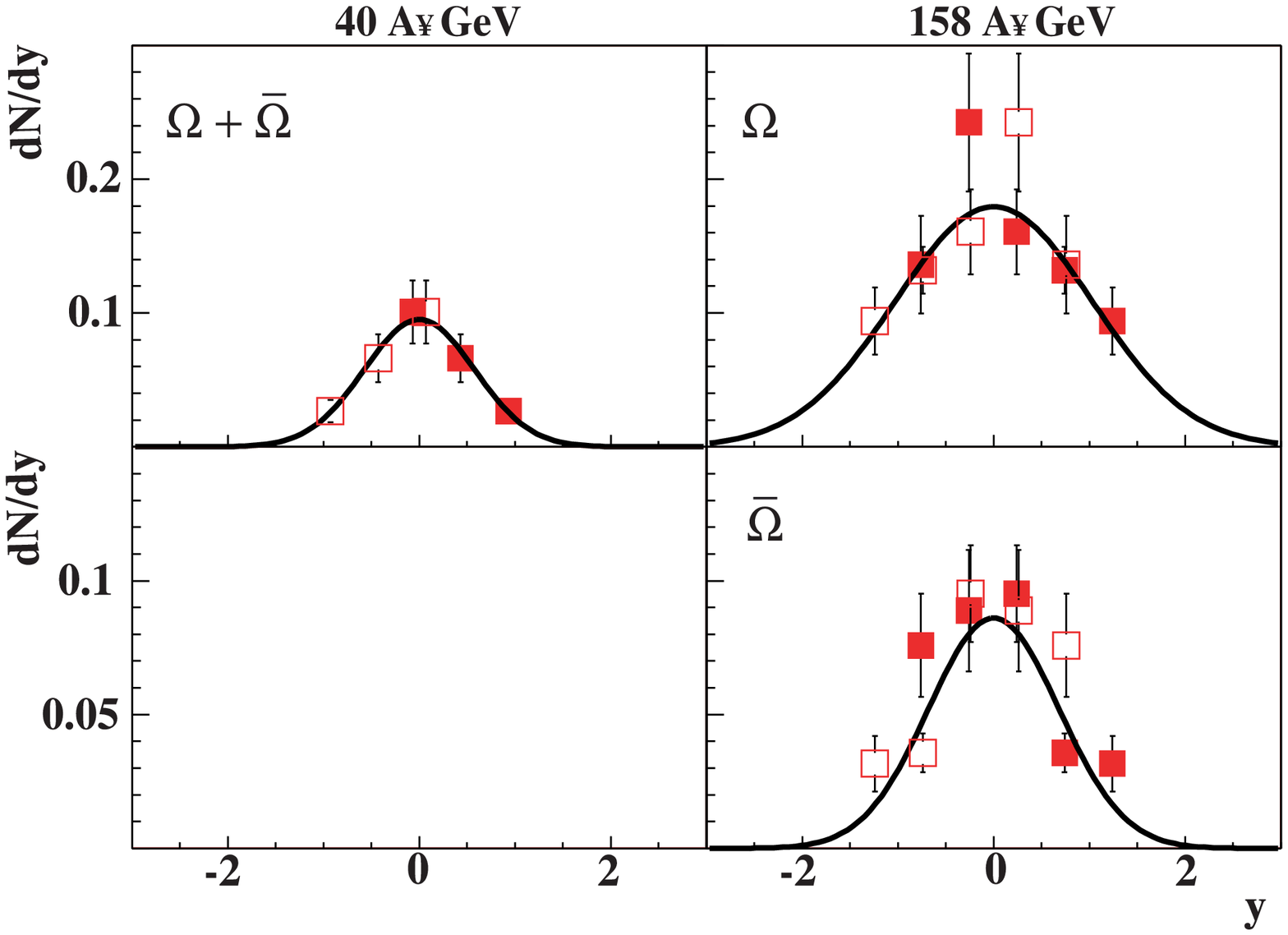} 
\caption{The rapidity spectra of $\Omega$ and $\bar{\Omega}$ in central Pb+Pb collisions at 40 and 158 A$\cdot$GeV. The upper row shows the rapidity distribution of $\Omega$ + $\bar{\Omega}$ at 40 A$\cdot$GeV and $\Omega$ at  158 A$\cdot$GeV and the lower row the $\bar{\Omega}$ rapidity distribution at 158 A$\cdot$GeV. The full symbols are the measured points and the open points shows their reflection with respect to midrapidity.}  
\label{fig:Rapidity_Spectra_Omega}
\end{center}
\end{figure}

The large acceptance of the NA49 experiment allows to measure hyperon spectra in a large rapidity interval. Figs.~\ref{fig:Rapidity_Spectra_Lambda} - \ref{fig:Rapidity_Spectra_Omega} show the distributions for all hyperons in central Pb+Pb collisions. \\
 \\
Where necessary an extrapolation was performed using reasonable parametrisations of the spectral shape. The $\Lambda$ spectra at 30 and 40 A$\cdot$GeV and the $\bar{\Lambda}$ spectra at 40 to 158 A$\cdot$GeV are fitted by a Gaussian. At 80 A$\cdot$GeV the $\Lambda$-spectra are parametrised by the sum of two Gaussians positioned symmetrically with respect to midrapidity. The $\Lambda$ data at 158 A$\cdot$GeV were extrapolated using a procedure described in \cite{A.6}. For all hyperons the mean multiplicities in the full phase-space were calculated by summing the measured points and extrapolating to full rapidity coverage using the shown parametrisation. It is seen that the shape of the $\Lambda$ rapidity distribution changes with the energy. The width of the distribution increases with energy (see table~\ref{tab:Lambda_Table}). The total yield is approximately independent of energy, but the midrapidity density decreases with energy. The $\bar{\Lambda}$ rapidity distribution also gets broader with energy. In this case, both the total yield and the midrapidity density increase with energy. \\
 \\
In Fig.~\ref{fig:Rapidity_Spectra_Xi} and~\ref{fig:Rapidity_Spectra_Omega} the rapidity distribution of $\Xi$ and $\Omega$ are shown. These show the same qualitative features and i.e. a broadering with energy and increase of both midrapidity and total yield. The numerical values of the Gaussian width, the total yield and the midrapidity density for the $\Lambda$, $\Xi$ and $\Omega$ hyperons are given in Tables~\ref{tab:Lambda_Table} - \ref{tab:Omega_Table}. 
\begin{table}[hbt!]
\begin{center}
\begin{tabular}{|c|c|c|c|c|}
& 30 A$\cdot$GeV & 40 A$\cdot$GeV & 80 A$\cdot$GeV & 158 A$\cdot$GeV \\[1pt]
\hline
T($\Lambda$) [MeV] & 245 $\pm$ 8 & 240 $\pm$ 6 & 252 $\pm$ 7 & 294 $\pm$ 10 \\
T($\bar{\Lambda}$) [MeV] &  & 300 $\pm$ 11 & 298 $\pm$ 13 & 270 $\pm$ 7 \\ 
dN/dy ($\Lambda$) & 16.5 $\pm$ 0.2 & 15.3 $\pm$ 0.6 & 13.5 $\pm$ 0.7 & 11.4 $\pm$ 1.0 \\ 
dN/dy ($\bar{\Lambda}$) &  & 0.42 $\pm$ 0.04 & 1.06 $\pm$ 0.08 & 1.69 $\pm$ 0.17 \\ 
$\sigma$ ($\Lambda$) & 1.0 $\pm$ 0.2 & 1.16 $\pm$ 0.1 & &   \\ 
$\sigma$ ($\bar{\Lambda}$) &  & 0.7 $\pm$ 0.1 & 0.9 $\pm$ 0.1 & 1.0 $\pm$ 0.2 \\ 
$\langle \Lambda \rangle$ & 41.9 $\pm$ 3.2 & 45.6 $\pm$ 1.9 & 47.4 $\pm$ 2.8 & 44.1 $\pm$ 3.2 \\ 
$\langle \bar{\Lambda} \rangle$ & & 0.74 $\pm$ 0.04 & 2.3 $\pm$ 0.3 & 3.9 $\pm$ 0.2 \\ \hline
\end{tabular}
\caption{The inverse slope parameter $T$ of the transverse mass spectra (fitted in the range $m_{t}$-$m_{0}$ $>$ 0.25 GeV/$c^{2}$) at midrapidity, the midrapidity density dN/dy, the Gaussian width $\sigma$ of the rapidity distribution and the total multiplicity $\langle N \rangle$ for $\Lambda$ and $\bar{\Lambda}$ in central Pb+Pb collisions at 30, 40, 80 and 158 A$\cdot$GeV.} 
\label{tab:Lambda_Table} \vspace{0.2cm}
\begin{tabular}{|c|c|c|c|}
 & $\Xi$ & $\Xi$ & $\bar{\Xi}$ \\
 & 40 A$\cdot$GeV  & 158 A$\cdot$GeV & 158 A$\cdot$GeV\\[1pt]
\hline
T [MeV] & 210 $\pm$ 11 & 261 $\pm$ 6 & 292 $\pm$ 11 \\
dN/dy & 1.07 $\pm$ 0.06 & 1.49 $\pm$ 0.08 & 0.33 $\pm$ 0.04 \\ 
$\sigma$ &  0.88 $\pm$ 0.07 & 1.17 $\pm$ 0.07 &  0.87 $\pm$ 0.07  \\ 
$\langle N \rangle$ & 2.41 $\pm$ 0.15 & 4.12 $\pm$ 0.20 & 0.77 $\pm$ 0.04 \\ \hline
\end{tabular}
\caption{The inverse slope parameter $T$ of the transverse mass spectra (fitted in the range $m_{t}$-$m_{0}$ $>$ 0.25 GeV/$c^{2}$) at midrapidity, the midrapidity density dN/dy, the Gaussian width $\sigma$ of the rapidity distribution and the total multiplicity $\langle N \rangle$ for $\Xi$ and $\bar{\Xi}$ in central Pb+Pb collisions at 40 and 158 A$\cdot$GeV.} \vspace{0.2cm}
\label{tab:Xi_Table} 
\begin{tabular}{|c|c|c|c|}
 & $\Omega+\bar{\Omega}$ & $\Omega$ & $\bar{\Omega}$ \\
 & 40 A$\cdot$GeV  & 158 A$\cdot$GeV & 158 A$\cdot$GeV\\[1pt]
\hline
T [MeV] & 218 $\pm$ 39 & 276 $\pm$ 23 & 285 $\pm$ 38 \\
dN/dy & 0.10 $\pm$ 0.02 & 0.24 $\pm$ 0.04 & 0.11 $\pm$ 0.02 \\ 
$\sigma$ &  0.6 $\pm$ 0.1 & 1.0 $\pm$ 0.2 &  0.7 $\pm$ 0.1  \\ 
$\langle N \rangle$ & 0.20 $\pm$ 0.03 & 0.47 $\pm$ 0.07 & 0.15 $\pm$ 0.02 \\ \hline
\end{tabular}
\caption{The inverse slope parameter $T$ of the transverse mass spectra (fitted in the range $m_{t}$-$m_{0}$ $>$ 0.25 GeV/$c^{2}$) at midrapidity, the midrapidity density dN/dy, the Gaussian width $\sigma$ of the rapidity distribution and the total multiplicity $\langle N \rangle$ for $\Omega$ and $\bar{\Omega}$ in central Pb+Pb collisions at 40 and 158 A$\cdot$GeV.} 
\label{tab:Omega_Table}
\end{center}
\end{table} 
 
\section{Energy dependence of hyperon production}

\begin{figure}[hbt!]
\begin{center}
\includegraphics[height=9.5cm, width=9.5cm]{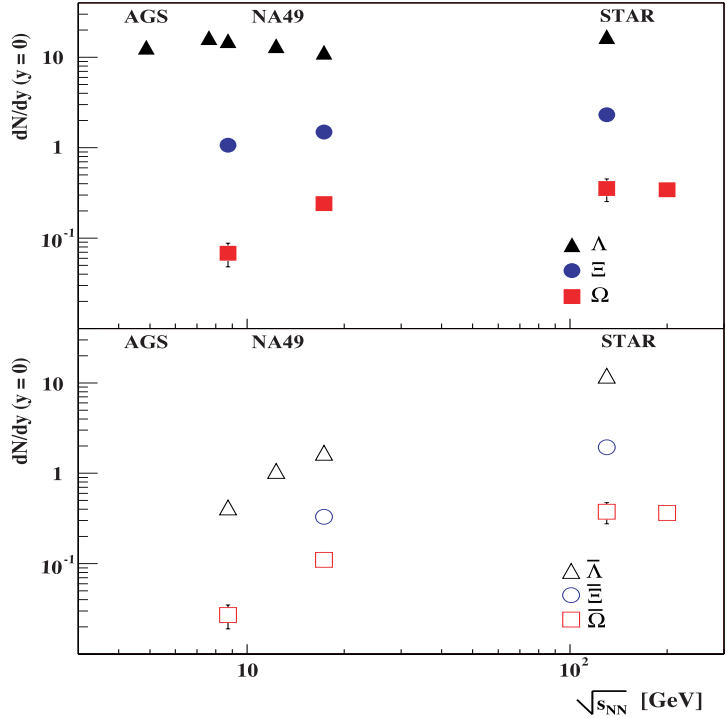} 
\caption{Hyperon (top) and antihyperon (bottom) rapidity density at midrapidity from AGS to RHIC energies.}  
\label{fig:midrapidity_yields}
\end{center}
\end{figure}

\begin{figure}[hbt!]
\begin{center}
\includegraphics[scale=0.5]{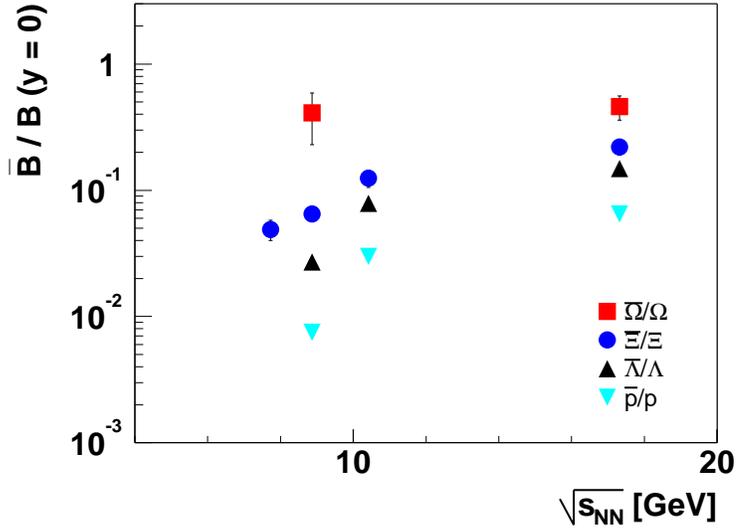}  
\caption{The antibaryon/baryon ratio ($\bar{B}$$/$$B$) at midrapidity in the SPS energy range.}  
\label{fig:Baryon_ratio}
\end{center}
\end{figure}

\begin{figure}[hbt!]
\begin{center}
\includegraphics[scale=0.6]{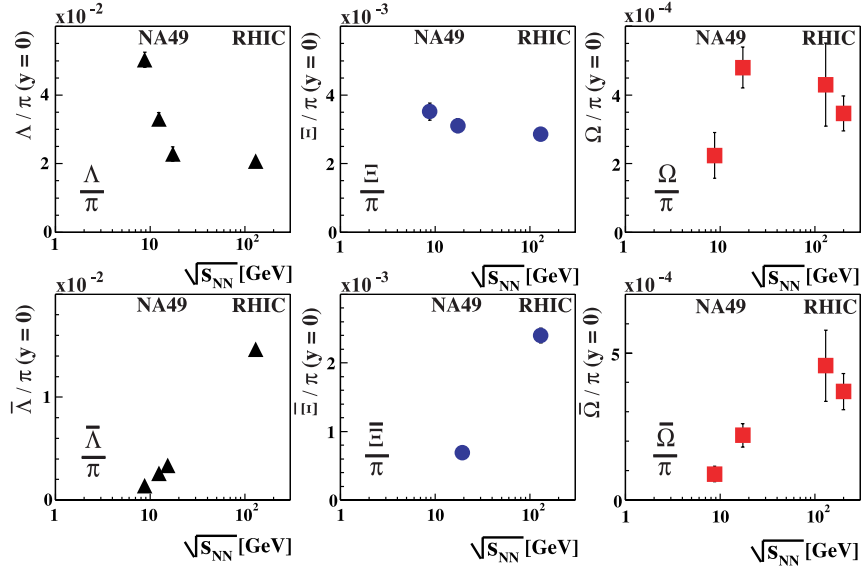} 
\caption{Energy dependence of the midrapidity hyperon/pion (top) and antihyperon/pion (bottom) ratio in central Pb+Pb and Au+Au collisions.}  
\label{fig:Hyperon_Pion_ratio_midrapidity}
\end{center}
\end{figure}

\begin{figure}[hbt!]
\begin{center}
\includegraphics[scale=0.65]{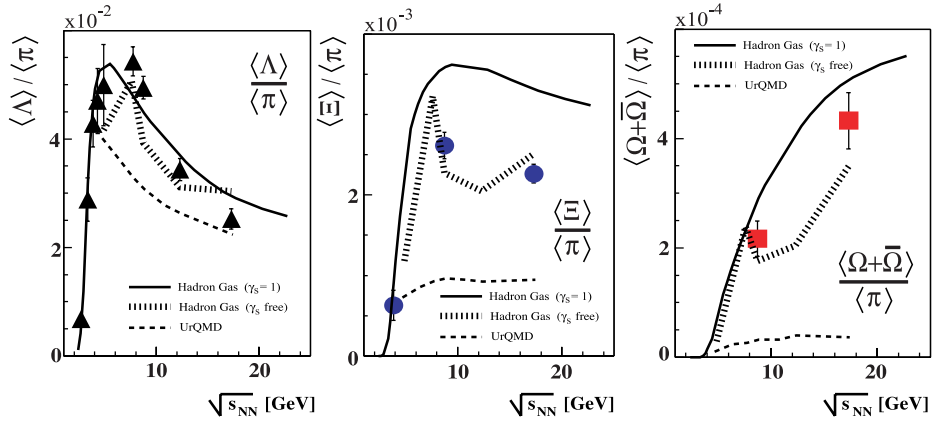}  
\caption{Energy dependence of the hyperon/pion ratio (total yields) in central Pb+Pb collisions compared to model predictions.}  
\label{fig:Hyperon_Pion_ratio_total}
\end{center}
\end{figure}

NA49 measured hyperon production in central Pb+Pb collisions from 30 to 158 A$\cdot$GeV. The STAR Collaboration at RHIC published hyperon densities at midrapidity in central Au+Au collisions at $\sqrt{s_{NN}}$ = 130 GeV and  $\sqrt{s_{NN}}$ = 200 GeV \cite{A.9} - \cite{A.11}. Fig.~\ref{fig:midrapidity_yields} shows the energy dependence of hyperon midrapidity density. The $\Lambda$ midrapidity density increases from AGS \cite{A.12} to SPS energies. At SPS energies the $\Lambda$ midrapidity density decreases from low to top SPS energy and it increases again from the top SPS to RHIC energies. The $\bar{\Lambda}$ midrapidity density shows a very different trend, it increases from the low SPS to RHIC energy. The $\Xi$ and $\Omega$ and their antiparticles also show an increase with energy. \\ \\
The NA49 antibaryon/baryon$^{b}$ ratios are shown in Fig.~\ref{fig:Baryon_ratio}  as a function of the beam energy. The ratios increase with the increasing strangeness content of the hyperons at each energy. The energy dependence of the ratio decreases with the strangeness content of the particle. This general trend can be understood as a result of a decrease of the baryon density at midrapidity with increasing energy. \\ \\
 \\
In Fig.~\ref{fig:Hyperon_Pion_ratio_midrapidity} (top) the energy dependence of the hyperon density at midrapidity normalised by pion ($\pi$ = 1.5 ($\pi^{-}$ + $\pi^{+}$)) are shown. The ratio decreases from low to top SPS energies. The STAR data at $\sqrt{s_{NN}}$ = 130 GeV are somewhat lower than the point at the top SPS energy. The maximum seems to disappear with the increasing strangeness of the particle. The $\Omega$/$\pi$ ratio shows a different trend. The ratio increases from low to top SPS energies and it seems to be constant from the top SPS to higher energies. The antihyperon/$\pi$ ratios at midrapidity show a monotonic increase from SPS to RHIC energies. \\ \\ 
The ratio of the total yield of $\Lambda$, $\Xi$ and $\Omega$ to pions ($\langle \pi \rangle$ = 1.5 ($\langle \pi^{-} \rangle$ + $\langle \pi^{+} \rangle$)) is shown as a function of $\sqrt{s_{NN}}$ in Fig.~\ref{fig:Hyperon_Pion_ratio_total}. The $\langle \Lambda \rangle$/$\langle \pi \rangle$ ratio shows a maximum located at about 30 A$\cdot$GeV. The maximum again disappear with the increasing strangeness of the particle.  \\   
 \\
The $\langle \Omega + \bar{\Omega} \rangle$/$\langle \pi \rangle$ ratio increases from low to top SPS energies. The dotted lines show predictions of the string hadronic model UrQMD~\cite{A.13}. This model generally underpredicts the measurements and the strongest disagreement (factor 6-10) is seen for $\Omega$s. The solid line shows a prediction of a hadron gas model with the strangeness saturation factor $\gamma_{s}$ = 1 \cite{A.14}. This model approximately (within 20 \%) reproduces the data for $\langle \Lambda \rangle$/$\langle \pi \rangle$ and $\langle \Omega + \bar{\Omega} \rangle$/$\langle \pi \rangle$ ratios. The $\langle \Xi \rangle$/$\langle \pi \rangle$ ratio is overpredicted by about 40 \%. On the other hand the hadron gas model in which the strangeness saturation factor is allowed to be free reproduces all ratios quite well~\cite{A.15}. 

\section{Conclusions}

New NA49 results on hyperon production in Pb+Pb interactions at CERN SPS energies were presented. The transverse mass spectra, the rapidity distributions, midrapidity density and total yields are shown and their energy dependence is discussed. The results are compared to models. Rich data might shed more light on strangeness production and hadronisation.

\section*{Acknowledgements}
This work was supported by the US Department of Energy
Grant DE-FG03-97ER41020/A000,
the Bundesministerium fur Bildung und Forschung, Germany, 
the Polish State Committee for Scientific Research (2 P03B 130 23, SPB/CERN/P-03/Dz 446/2002-2004, 2 P03B 02418, 2 P03B 04123), 
the Hungarian Scientific Research Foundation (T032648, T032293, T043514),
the Hungarian National Science Foundation, OTKA, (F034707),
and the Polish-German Foundation. 

\section*{Notes}  
\begin{notes}
\item[a]
For a full author list of the NA49 Collaboration see \cite{A.0}
\item[b]
The ratios of $\Xi$ at 30, 40 and 80 A$\cdot$GeV are uncorrected for acceptance and efficiency
\end{notes}

\vfill\eject
\end{document}